\documentclass[twocolumn,aps,prb]{revtex4} 

\usepackage{graphics}

\begin{document}


\bibliographystyle{apsrev}

\title{Understanding Popper's experiment}

\author{Tabish Qureshi}
\email{tabish@jamia-physics.net}
\affiliation{Department of Physics, Jamia Millia Islamia, New Delhi-110025,
India}

\begin{abstract}
An experiment proposed by Karl Popper is considered by many to be a crucial
test of quantum mechanics. Although many loopholes in the original proposal
have been pointed out, they are not crucial to the test. We use only the
standard interpretation of quantum mechanics to point out what is
fundamentally wrong with the proposal, and demonstrate that Popper's basic
premise was faulty.
\end{abstract}


\maketitle

\section{Introduction}

Quantum theory is a tremendously successful theory when it comes to explaining
or predicting physical phenomena. However, there is no consensus on how it
is to be interpreted. 
For example, it is not clear whether the wave function
is to be considered a real object or just a mathematical tool for calculating
probabilities. However, these debates do not seem to have any bearing on the
predictions for the outcomes of experiments based on quantum theory. Thus,
most scientists continue to use quantum mechanics as a tool, leaving the
debate on its meaning to others.

Karl Popper, a philosopher of science, has proposed an experiment to test the
standard interpretation of quantum theory.\cite{popper,popper1} Popper's
experiment is of much interest because the outcome depends on the
interpretation of quantum
theory.\cite{sudbery,sudbery2,krips,collet,storey,redhead,plaga,short,nha,peres,hunter}
Ideas that used to fall under the realm of philosophy
appeared to be testable. New interest was generated by its experimental
realization by Kim and Shih\cite{shih} and by claims that it proved the
absence of quantum nonlocality.\cite{unni} At the heart of Popper's proposal
is the concept of entanglement, which is a unusual quantum phenomenon.
Spatially separated, entangled particles, seem to depend on each other, even
though there is no physical interaction. The implications of entangled
states were discussed by Einstein, Podolsky and Rosen (EPR) in their famous
paper.\cite{epr} Such states are now commonly referred to as EPR states.

\section{Popper's Proposed Experiment}
Popper's proposed experiment consists of a source $S$ that can generate pairs
of particles traveling to the left and to the right along the $x$-axis. The
momentum along the $y$-direction of the two particles is entangled in such a
way so as to conserve the initial momentum at the source,
which is zero. There are two slits, one each in the paths of the two particles.
Behind the slits are semicircular arrays of detectors which can detect the
particles after they pass through the slits (see Fig.~1).
\begin{figure}[h]
\resizebox{3.5in}{!}{\includegraphics{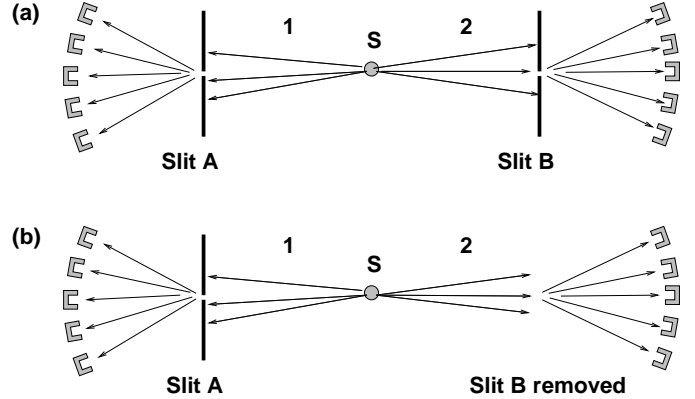}}
\caption{Schematic diagram of Popper's thought experiment. (a) With both 
slits, the particles are expected to show scatter in momentum. (b) By removing
slit B, Popper believed that the standard interpretation of
quantum mechanics could be tested.}
\end{figure}

Being entangled in momentum space implies that in the absence of the two
slits, if a particle on the left is measured to have a momentum $p$, the particle
on the right will necessarily be found to have a momentum $-p$. One can
imagine a state similar to the EPR state, \cite{epr}
$\psi(y_1,y_2) = \!\int_{-\infty}^{\infty}e^{ipy_1/\hbar} e^{-ipy_2/\hbar}dp$.
As we can see, this state also implies that if a particle on the left is
detected at a distance $y$ from the horizontal line, the particle on the right
will necessarily be found at the same distance $y$ from the horizontal line.
A tacit assumption in Popper's setup is that the initial spread in momentum
of the two particles is not very large.
Popper argued that because the slits localize the particles to a narrow
region along the $y$-axis,
they experience large uncertainties in the $y$-components of their momenta.
This larger spread in the momentum will show up as particles being
detected even at positions that lie outside the regions where particles
would normally reach based on their initial momentum spread.
The momentum spread, because of a real slit, is expected. 
 
Popper suggested that slit B be made very large (in effect, removed).
In this situation, Popper argued
that when particle 1 passes through slit A, it is localized to
within the width of the slit. 
He further argued that the standard interpretation of quantum mechanics
tells us that if particle 1 is localized in a small region of space, particle
2 should become similarly
localized, because of entanglement.
In fact, when this experiment is done without the slits, the correlation
in the detected positions of particles 1 and 2, is an example of such a
localization. 
Popper completed his argument by saying that if particle 2 is
localized in a narrow region of space, its momentum
spread also will increase, causing other detectors to register:
\begin{quote}
``We thus obtain fairly precise {\em `knowledge'} about $y(B)$ -- we have
'measured' it indirectly. And since it is, according to the Copenhagen
interpretation, our {\em knowledge} which is described by the theory ---
and especially by the Heisenberg relations --- we should expect that the
momentum \ldots\ of the beam that passes through slit B scatters as much as
that of the beam that passes through slit A, even though the slit A is
much narrower \ldots\ If the Copenhagen interpretation is correct, then such
counters on the far side of slit B that are indicative of a wide scatter
\ldots should now count coincidences; counters that did not count any
particles before the slit A was narrowed \ldots''\cite{popper}
\end{quote}

Popper had reasons to believe that if one were to actually carry out the
experiment, particle 2 would not show any additional momentum spread. He
argued that this absence of additional momentum spread 
would prove that the standard interpretation of quantum mechanics was wrong.

Popper believed that quantum mechanics could be interpreted ``realistically,''
so that we could talk of the position and momentum of a particle at the same
time. He also did not like the notion, which is central to the Copenhagen
interpretation of quantum mechanics, that the knowledge gained about particle
1 could have any influence on particle 2. Thus, he intended to demonstrate by
this experiment that a position measurement on particle 1 would have no
effect on the momentum spread of particle 2.

\section{Objections to Popper's Experiment}

In 1985, Sudbery pointed out that the EPR state already contained an infinite
spread in momenta, so no further spread could be seen by localizing one
particle. \cite{sudbery,sudbery2} Sudbery further stated that collimating the original
beam, so as to reduce the momentum spread, would destroy the correlations
between particles 1 and 2. We will show that having a reduced momentum
spread doesn't completely destroy the correlations. The presence of correlations
despite a reduced momentum spread, 
is also seen in the experimentally observed spontaneous parametric
down-conversion (SPDC) photon pairs.\cite{strekalov}

In 1987 there came a major objection to Popper's proposal from Collet and
Loudon. \cite{collet} They pointed out that because the particle pairs
originating from the source had a zero total momentum, the source could not
have a sharply defined position. They showed that once 
the uncertainty in the position of the source is taken into account, the
blurring introduced washes out the Popper effect. However, it has been
demonstrated that a point source is not crucial for Popper's experiment, and
a broad SPDC source can be set up to give a strong correlation between the
photon pairs.\cite{strekalov}

Redhead analyzed Popper's experiment with a broad source and concluded that
it could not yield the effect Popper that was seeking.\cite{redhead} However,
a modified setup using a broad source and a converging lens has been shown to
lead to a localizing effect.\cite{strekalov}

Popper's experiment was realized in 1999 by Kim and Shih using a SPDC photon
source.\cite{shih} They did not observe an extra spread in the
momentum of particle 2 due to particle 1 passing through a narrow slit. In
fact, the observed momentum spread was narrower than that contained in the
original beam. This observation seemed to imply that Popper was right. Short
has criticized Kim and Shih's experiment, arguing that because of the finite
size of the source, the localization of particle 2 is imperfect,\cite{short}
which leads to a smaller momentum spread
than expected. However, Short's argument implies that if the source were
improved, we should see a spread in the momentum of particle 2.

We have analyzed Popper's proposal and showed that the mere presence of
slit A doesn't lead to a reduction of the wavefunction.\cite{tabish} So, we
should not expect any effect of slit A on particle 2. We concluded that in
the original Popper's proposal and in Kim and Shih's realization, we
would not see any spread in the momentum of particle 2, just due to the
presence of slit A in the path of particle 1. This conclusion was based on
the standard interpretation of quantum mechanics. Our conclusion also implied
that even if the source is improved to give a better correlation of photons,
we would not see any spread in the momentum of particle 2. So, Popper may
have been right in saying that there would be no spread, but for the wrong
reasons.

\section{What is wrong with Popper's proposal?}
It is easy to see that our earlier objection to Popper's experiment can be
remedied
by putting a detector immediately behind slit A, such that a photon passing
through the slit is detected immediately. In this case, as soon as the
particle passes through slit A, we acquire the information that causes a
reduction of the wavefunction because of the detector. The question we 
now ask is will we see any extra spread in the momentum
of particle 2? After all, we can make the slit A as narrow as we want, and
the resultant localization of particle 2 should lead to an increasing
momentum spread.

Let us investigate this scenario rigorously. From practical
considerations,
the initial momentum spread has to be finite. Let us assume an initial
wavefunction of the form:
\begin{equation}
\psi(y_1,y_2) = A\!\int_{-\infty}^\infty dp
e^{-p^2/4\sigma^2}e^{-ipy_2/\hbar} e^{i py_1/\hbar}
\exp[-{(y_1+y_2)^2\over 16\Omega_0^2}], \label{state}
\end{equation}
where $A$ is a constant necessary for the normalization of $\psi$. Without the
$e^{-(y_1+y_2)^2/16\Omega_0^2}$ term, the state (\ref{state}) would be
infinitely extended.

We have neglected the spread in the wavefunction as the particles travel
to reach the position of the slits. Eq. (\ref{state}) represents
the wavefunction of the particles at the instant when particle 1 reaches slit A.
Motion along the
$x$-axis is not very interesting as far as entanglement is concerned, and
thus has been ignored here. Although Eq.~(\ref{state}) represents an
entangled state, where the individual states of particles 1 and 2 have no
meaning, we can still talk about uncertainty in any variable we 
choose. We define the uncertainty in a variable $Q$ as
\begin{equation}
	\Delta Q = \sqrt{ \langle\psi| (\hat{Q} 
 -\langle\psi|\hat{Q}|\psi\rangle)^2 |\psi\rangle },
\end{equation}
where $|\psi\rangle$ could be an entangled state. We first calculate
the uncertainty in the momentum of, say, particle 2. The wavefunction defined
by Eq.~(\ref{state}) after integrating over $p$, also can be written as:
\begin{equation}
\psi(y_1,y_2) = 2A \sqrt{\pi}\sigma e^{-(y_1-y_2)^2\sigma^2/\hbar^2}
e^{-(y_1+y_2)^2/16\Omega_0^2} .
\label{newstate}
\end{equation}
Because $\langle\hat{p}_{2y}\rangle = 0$, the uncertainty in $p_{2y}$ is
given by
\begin{eqnarray}
\Delta p _{2y} &=& \Big[|A|^2 4\pi\sigma^2 \!\int_{-\infty}^\infty dy_1
\!\int_{-\infty}^\infty dy_2\, e^{-{(y_1-y_2)^2\sigma^2\over\hbar^2}} e^{-
{(y_1+y_2)^2\over 16\Omega_0^2}}\nonumber\\
&&\big(-\hbar^2{d^2\over
dy_2^2}\big) e^{-{(y_1-y_2)^2\sigma^2\over\hbar^2}} e^{- {(y_1+y_2)^2\over
16\Omega_0^2}} \Big]^{1/2}\nonumber\\
	&=& [\sigma^2 + {\hbar^2\over 16\Omega_0^2}]^{1/2}. \label{dp2}
\end{eqnarray}
Because the state (\ref{newstate}) is symmetric in $y_1$ and $y_2$, the
uncertainty in $p_{1y}$ 
is also the same as that for $p_{1y}$. The position uncertainty of the
two particles is
$\Delta y_1 =
\Delta y_2 =
\sqrt{\Omega_0^2+\hbar^2/16\sigma^2}$.

Let us suppose that a measurement is performed on particle 1 at slit A
such that the wavefunction of particle 1 is reduced to
\begin{equation}
\phi_1(y_1) = \frac{1}{(\epsilon^22\pi)^{1/4} } e^{-y_1^2/4\epsilon^2}.
\end{equation}
In this state, the uncertainty in $y_1$ is given by
\begin{equation}
	\Delta y_1 = \sqrt{\langle\phi_1|(\hat{y}_1 - \langle\hat{y}_1\rangle
 )^2|\phi_1\rangle} 
= \epsilon.
\end{equation}

After the measurement, the particles are disentangled, and the subsequent
evolution of one is independent of the other in the sense that they are
governed by different wavefunctions.
The wavefunction of particle 2 is now reduced to:
\begin{eqnarray}
\phi_2(y_2) &=& \!\int_{-\infty}^\infty \psi(y_1,y_2) \phi_1^*(y_1)
dy_1\nonumber= {2A\sqrt{\pi}\sigma\over (\epsilon^22\pi)^{1/4}
}
\!\int_{-\infty}^\infty 
 e^{-{(y_1-y_2)^2\sigma^2\over\hbar^2}} e^{-{(y_1+y_2)^2\over 16\Omega_0^2}}
 e^{-{y_1^2\over 4\epsilon^2}}dy_1\nonumber\\
&=& {2A\sqrt{\pi}\sigma\over (\epsilon^22\pi)^{1/4}
}\sqrt{\pi\alpha}
\,e^{-y_2^2/4\Omega^2}, \label{phi2}
\end{eqnarray}
where $\alpha={\sigma^2\over\hbar^2}+{1\over 16\Omega_0^2}+{1\over
4\epsilon^2}$, and
\begin{equation}
\Omega = \sqrt{\frac{\epsilon^2(1+{\hbar^2\over 16\sigma^2\Omega_0^2})+\hbar^2/4\sigma^2}{1+{\epsilon^2\over\Omega_0^2}
 + {\hbar^2\over 16\sigma^2\Omega_0^2}}}.
\end{equation}

{}From Eq.~(\ref{phi2}) it follows that the uncertainty in the position of
particle 2 is given by:
\begin{equation}
\Delta y_2 = \sqrt{\frac{\epsilon^2(1+{\hbar^2\over 16\sigma^2\Omega_0^2})+\hbar^2/4\sigma^2}{1+{\epsilon^2\over\Omega_0^2}
 + {\hbar^2\over 16\sigma^2\Omega_0^2}}} .\label{dy2}
\end{equation}
Equation~(\ref{dy2}) implies that when a measurement is performed on particle
1, so as to localize it within a spatial region $\epsilon$, particle 2 becomes
localized in a region $\Delta y_2$ given by Eq.~(\ref{dy2}). Once particle 2
is localized to a narrow region in space, its subsequent evolution should
show the momentum spread dictated by the uncertainty principle.
The uncertainty in the momentum of particle 2 is now given by
\begin{equation}
\Delta p_{2y} = {\hbar\over 2\Delta y_2}
 = \sqrt{\frac{\sigma^2(1+\epsilon^2/\Omega_0^2)+
 \hbar^2/16\Omega_0^2}{1+4\epsilon^2(\sigma^2/\hbar^2+1/16\Omega_0^2)}}.
\label{dp2n}
\end{equation}

Now we have all the results needed to examine what happens in Popper's
experiment. Let us look for the maximum possible scatter in the momentum of particle
2. To do so we have to localize particle 1 in a very narrow region, which is
what Popper wanted to achieve by narrowing slit A. Let us look at the momentum
uncertainty of particle 2 in the limit $\epsilon\rightarrow 0$:
\begin{equation}
\label{eq:dp2}
\lim_{\epsilon\to 0} \Delta p_{2y} = \sqrt{\sigma^2+
 \hbar^2/16\Omega_0^2}.
\end{equation}
But the right-hand side of Eq.~(\ref{eq:dp2}) is exactly the uncertainty in
the momentum of particle 2 in the initial state (\ref{state}), before particle
1 entered the slit (see
Eq.~(\ref{dp2})). So, even in the
best case, there is no extra spread in the momentum of particle 2. In fact,
we can show that the momentum spread described by Eq.~(\ref{dp2n}) is less
than or equal to that described by Eq.~(\ref{dp2}) for any value of
$\epsilon$,
$\sigma$, and $\Omega_0$. This fact, that there is no extra momentum spread
in particle 2, is at variance with what
Popper had concluded regarding the standard interpretation of quantum
mechanics. On the other hand, the momentum spread of particle 1 after the
measurement is given by
\begin{equation}
\Delta p_{1y} = {\hbar\over 2\Delta y_1} 
= {\hbar\over 2\epsilon},
\end{equation}
which, for $\epsilon\to 0$ will become infinite.

We note that if $\epsilon<<\Omega_0$, the position
spread of particle 2 becomes smaller as a result of the measurement
performed on particle 1. However, the momentum spread of particle 2 as
given in Eq.~(\ref{dp2n}) also is smaller than the original spread given by
Eq.~(\ref{dp2}). This smaller momentum spread is possible because the
original state is not a minimum uncertainty state. The spread in both 
conjugate variables can thus be reduced at the same time, within limits of
course. Because the momentum of particle 2 cannot show any additional spread
for a minimum uncertainty initial state, the position spread of particle 2
also should remain unchanged during the measurement performed on particle
1. Indeed, a calculation confirms an interesting scenario. For
$\Omega_0={\hbar\over 4\sigma}$, the initial state is a minimum uncertainty
state for particles 1 and 2. With this choice of $\Omega_0$, Eqs.~(\ref{dp2})
and (\ref{dp2n}) are identical and equal to $\sqrt{2}\sigma$. In this case,
the initial position spread of particle 2 is $\Delta y_2 =
\hbar/(2\sqrt{2}\sigma)$, which is the identical result given by
Eq.~(\ref{dy2}). This result implies that a position measurement on
particle 1 has absolutely no effect on particle 2. This conclusion
might look very surprising, but we can check that the initial state
(\ref{newstate}) becomes disentangled for 
$\Omega_0=\hbar/4\sigma$.

We see that the fundamental mistake made by Popper was to assume that
according to the standard interpretation and a finite initial momentum
spread, localizing particle 1 to a narrow region will lead to the
localization of particle 2 in a region as narrow. In contrast, we have seen
that if particle 1 is localized to a region of size
$\epsilon$, particle 2 is localized to
$\sqrt{\frac{\epsilon^2(1+\hbar^2/16\Omega_0^2\sigma^2)+\hbar^2/4\sigma^2}{1+\epsilon^2/\Omega_0^2
 + \hbar^2/16\sigma^2\Omega_0^2}}$. Only in the limits $\sigma\to\infty$ and
$\Omega_0\to\infty$, does the latter reduce to $\epsilon$. But in that case,
the initial momentum spread is already infinite. Short's argument regarding
the finite size source leading to imperfect localization of the second
particle\cite{short} is fully consistent with the general analysis presented
here.

Finally, we verify if quantum mechanics shows what, in Popper's view,
would constitute an effect of the position measurement of particle 1 on
particle 2. Popper states: ``To sum up: if the Copenhagen interpretation is
correct, then any increase in the precision in the measurement of our {\em
mere knowledge} of the particles going through slit B should increase their
scatter.''\cite{popper}
This view is less stringent -- it does not demand that the localization of
particle 2 be as much as that of particle 1. It just says that if the 
(indirect) localization of particle 2 is made more precise, the momentum
spread should show an increase. The momentum spread of
particle 2 in Eq.~(\ref{dp2n}) in the limit in which the correlation between
the two particles is expected to be stronger, namely
$\sigma\gg\hbar/4\Omega_0$, $\epsilon/\Omega_0\ll 1$, is 
\begin{equation}
\Delta p_{2y} \approx {\hbar\over \sqrt{\hbar^2/\sigma^2 + 4\epsilon^2}}.
\end{equation}
Clearly, if $\epsilon$ is decreased, $\Delta p_{2y}$ increases.
While analyzing Popper's experiment, Krips had predicted
that narrowing slit A would lead to momentum spread increasing at slit B,
which is the same as our conclusion. \cite{krips} Our result relies only on the
mathematics of quantum mechanics. So, we conclude that Krips's prediction was
correct. Krips also had correctly argued that this conclusion can be
justified using the formalism of quantum theory, independent of any
particular interpretation.\cite{krips}

We deduce that in a real experiment, for a general
initial state, the approximate
position localization of particle 1 would lead to a somewhat reduced momentum
spread of particle 2. This conclusion is in contradiction with what Popper
thought the Copenhagen interpretation implies and what many defenders of the
Copenhagen interpretation probably imagined. The measurement on particle
1 does have an influence on particle 2, although not of the form one might
had naively expected. So, there is no escape from quantum nonlocality. If
Popper had imagined that the Copenhagen interpretation implies that a
measurement on particle 1 would lead to additional scatter in the momentum of
the second particle, his discomfort with it was justified. Quantum mechanics
doesn't have that kind of nonlocal influence -- the nonlocality is only at
the level of correlations. The lesson is that quantum mechanics is full of
surprises, and we should be careful when analyzing thought experiments in
quantum physics.

\begin{acknowledgments}
The author thanks C.\ S.\ Unnikrishnan and Thomas Angelidis for useful
discussions and Tony Sudbery for valuable suggestions. The author also
wishes to thank the Centre for Philosophy and Foundations of
Science, New Delhi, for organizing annual meetings to promote
the exchange of ideas on the foundations of quantum mechanics.
\end{acknowledgments}

\end{document}